\documentclass[12pt,reqno,oneside]{amsart}
\usepackage{amssymb}

\def\strutdepth{\dp\strutbox}
\def\margnote#1{\strut\vadjust{\kern-\strutdepth\specialstar{#1}}}
\def\specialstar#1{\vtop to \strutdepth{
   \baselineskip\strutdepth
   \vss\llap{\footnotesize #1\hskip 1em}\null}}

\newtheorem{theorem}{Theorem}
\newtheorem{proposition}[theorem]{Proposition}
\newtheorem{corollary}[theorem]{Corollary}
\newtheorem{lemma}[theorem]{Lemma}

\theoremstyle{definition}

\newcommand\lp{\left(}
\newcommand\rp{\right)}

\newcommand\numero{$\text{n}^{\text{o}}$ }

\newcommand\I{\mathrm{i}}
\newcommand\E{\mathrm{e}}
\newcommand\trig{\mathop{\mathrm{trig}}}
\newcommand\supp{\mathop{\mathrm{supp}}}

\newcommand\sgn{\mathop{\mathrm{sgn}}}

\newcommand\reals{\mathbb{R}}
\newcommand\cnums{\mathbb{C}}
\newcommand\integers{\mathbb{Z}}
\newcommand\natnums{\mathbb{N}}

\newcommand\ck[1]{\breve{#1}}
\newcommand\ckalpha{\ck{\alpha}}

\newcommand\Hop{\mathcal{H}}
\newcommand\tHop{\tilde{\Hop}}

\newcommand\trho{\tilde{\rho}}

\let\spacev\mathbf

\newcommand\Tsp{\spacev{T}}
\newcommand\Vsp{\spacev{V}}

\title{Algebraic Exact Solvability of trigonometric-type Hamiltonians associated
  to root systems.}  

\author{Niky Kamran}

\address{Dept. of Math.\\McGill U.\\Montreal, Canada\\ H3A 2K6}
\email{nkamran@math.mcgill.ca}

\author{R. Milson}
\address{Dept of Math.\\Dalhousie U.\\Halifax, Canada\\B3J 3J5}
\curraddr{McGill University\\Montreal}
\email{milson@math.mcgill.ca}

\subjclass{02.20.Qs, 03.65.Fd}

\keywords{Olshanetsky-Perelomov Hamiltonians, exact-solvability, root
  systems}

\begin{document}
\begin{abstract}
  In this article, we study and settle several structural questions concerning the
exact solvability of the Olshanetsky-Perelomov quantum Hamiltonians corresponding to
an arbitrary root system. We show that these operators can be written as linear
combinations of certain basic operators admitting infinite flags of invariant
subspaces, namely 
 the Laplacian and the logarithmic gradient of invariant factors of the
  Weyl denominator. The coefficients of the constituent linear
  combination become the coupling constants of the final model. We also demonstrate
the $L^2$ completeness of the
  eigenfunctions obtained by this procedure, and describe a straight-forward
  recursive procedure based on the
  Freudenthal multiplicity formula for constructing the eigenfunctions explicitly. 
\end{abstract}

\maketitle \pagestyle{myheadings}

\markboth{}{Exact solvability and root systems}

\section{Introduction}
The potentials first discovered by Calogero and Sutherland \cite{C,Su} and
subsequently generalized to arbitrary root systems by Olshanetsky and
Perelomov \cite{OP} play a central role in the theory of classical and
quantum completely integrable systems. One of the main themes of the
original work by Olshanetsky and Perelomov was to establish quantum
complete integrability, that is the existence of complete sets of
commuting operators. The actual eigenfunctions of the corresponding
Hamiltonians were discussed in numerous subsequent publications
\cite{PRZ, RT, Se,So}.

Our purpose in this paper is study and settle a certain number of
basic structural questions concerning the exact solvability of the
Olshanetsky-Perelomov Hamiltonians. In order to outline the main results of
our paper, we first need to give a precise definition of what we mean
by exact solvability. We will adopt a promising approach, which has
recently arisen in the framework of the theory of quasi-exactly
solvable potentials \cite{Sh, T1, GKO1, GKO2}, by defining a quantum
Hamiltonian $\Hop$ to be {\it algebraically exactly solvable} if one
can explicitly construct an ordered basis for the underlying Hilbert
space such that the corresponding flag of subspaces is $\Hop$
-invariant. In terms of this approach, the first step in the treatment
of an exactly solvable operator must be the construction of an
infinite flag of finite dimensional vector spaces ordered by
inclusion, the determination of a collection of basic operators that
preserve this flag, and the demonstration that the operator in
question is generated by the basic ones. The second step is to prove
the $L^2$ completeness in the underlying Hilbert space of this family
of subspaces.

In order to fit the Olshanetsky-Perelomov Hamiltonians of trigonometric type into this framework, we first recall that
these Hamiltonians are indexed by irreducible root systems, with the Calogero-Sutherland potentials corresponding to
type $A_{n}$ root systems. We thus consider the vector space of trigonometric functions which are invariant under the
Weyl group
$W$ of the given root system $R$. The partial order relation on dominant weights gives rise to a natural flag of
finite-dimensional subspaces of this infinite-dimensional vector space.  It is quite evident that the flag in question
is preserved by the ordinary, multi-dimensional Laplacian. Less evident is the fact that one can obtain other
flag-preserving operators by factoring the Weyl denominator
$$
A = \prod_{\alpha\in R^+} \E^{\alpha/2} - \E^{-\alpha/2}$$
into
factors corresponding to the various orbits of the Weyl group on $R$.
It turns out (see Proposition \ref{prop:invspace}) that the gradient
of the logarithm of each of the resulting factors also preserves the
flag in question.  More generally, one obtains other flag-preserving
second-order operators by taking linear combinations of the Laplacian
and of these gradients.  The Olshanetsky-Perelomov Hamiltonians are
then obtained by a ground-state conjugation.  This approach also sheds
light on the presence of multiple coupling constants in some of the
models; the number of coupling constants is precisely the number of
invariant factors of $A$, i.e. the number of Weyl group orbits in $R$,
or equivalently the number of distinct root lengths. We then show that
if all the coupling constants are positive, then the action of the
Hamiltonian on each subspace of the flag is diagonalizable.  This is
the first main result of our paper; it is given in Theorem 1.  The
second main result concerns the $L^2$ completeness of the resulting
eigenfunctions in the underlying Hilbert space of $L^2$ functions on
the alcove of the root system $R$.  It is also interesting to note
that if all the coupling constants are equal to 1, then one recovers a
second-order differential operator whose eigenfunctions are precisely
the characters of the corresponding simple Lie algebras.  Thus the
coupling constants can be regarded as parameters in a deformation of
the classical characters.  In the classical case, if one re-expresses
the gradient of $\log A$ in terms of a formal power series, one
obtains Freudenthal's recursion formula for the character
coefficients.  This trick also works for the deformed characters, and
leads to a recursion formula that allows one to straightforwardly
compute the eigenfunctions of the Olshanetsky-Perelomov Hamiltonians.
This result is presented in Section \ref{sect:recform}.

We should point out that the Weyl-invariant deformed characters which appear in the expressions of the eigenfunctions
of the Olshanetsky-Perelomov trigonometric Hamiltonians are related by a change of variables to the multivariate Jacobi
polynomials which have been investigated by Heckman and Opdam \cite{HO}. In particular, the analogue of the Freudenthal
multiplicity formula which is at the basis of the recursion formula we give in Proposition 19 for the eigenfunctions of
the Hamiltonians also appears in the context of their study . We should also mention the interesting recent
contributions of Brink, Turbiner and Wyllard \cite {Br} in the general effort aimed at understanding the exact
solvability for multidimensional systems in an algebraic context.

\section{Trigonometric-Type Potentials Associated to Root Systems.}

We first recall the abstract definition of the trigonometric
Olsha\-netsky-Perelomov Hamiltonians in terms of root systems.  Let
$\Vsp$ be a finite-dimensional real vector space endowed with a
positive-definite inner product $(u,v)\in\reals,\, u,v\in \Vsp$. We
use this inner product to identify $\Vsp$ with $\Vsp^*$. The induced
positive-definite inner product on $\Vsp^*$ will also be denoted by
$(\cdot,\cdot)$.  Let $\Delta: C^{\infty}(\Vsp; \reals)\to
C^{\infty}(\Vsp; \reals)$ and $\nabla: C^{\infty}(\Vsp; \reals)\to
\Gamma(\mathrm{T}\Vsp)$ denote the corresponding Laplace-Beltrami and
gradient operators.

For a non-zero $\alpha\in \Vsp^*$, we set
$\ckalpha=2\alpha/(\alpha,\alpha)$ and let $s_\alpha$ denote the 
reflection across the hyperplane orthogonal to $\alpha$:
$$s_\alpha(\beta)= \beta- (\ckalpha,\beta)\alpha,\quad \beta\in
\Vsp^*.$$

By a root system, we mean a finite, spanning subset $R$ of $\Vsp^*$
such that $0 \notin R$, $s_\alpha(R)\subset R$ for all $\alpha\in R$
and $(\ckalpha,\beta)\in \integers$ for all $\alpha,\beta\in R$.
\par\noindent
A root system $R$ is said to be irreducible if it cannot be
partitioned into a union of root systems spanning orthogonal subspaces
of $\Vsp$.

To any root system $R$ corresponds a root lattice $Q=\{\sum_R m_\alpha
\alpha : m_\alpha\in \integers\}$ and a weight lattice $P=\{\lambda
\in \Vsp^{*}:( \ckalpha,\lambda ) \in \integers \quad \forall
\alpha\in R\}$. The Weyl group of $R$, generated by $s_\alpha,\,
\alpha\in R$, will be denoted by $W$. The subgroup of $W$ fixing a
particular $\lambda \in \Vsp^*$ will be denoted by $W_{\lambda}$.

The hyperplanes $\{\lambda\in \Vsp^*: (\alpha,\lambda)=0\},\,\alpha\in
R$ define a set of open Weyl chambers in $\Vsp^*$. We choose a Weyl
chamber $C$ and let $R^+= R\cap \overline{C}$ denote the corresponding
subset of positive roots.  Let $B\subset R^+$ denote the set of simple
roots, i.e. the positive roots that cannot be written as the sum of
two positive roots.  Let $P^+=R\cap\overline{C}$ denote the set of
dominant weights.

We will say that a real number $c>0$ is a {\it  root length} if there exists
a $\alpha\in R$ such that $c=\|\alpha\|$.  Let $c$ be a root
length, and set
\begin{eqnarray*}
R_c &=& \{ \alpha\in R: \|\alpha\|=c\},\\
R_c^+ &=& R_c\cap R^+,\\
U_c &=& \frac{c^2}4 \sum_{\alpha\in R_c^+} \csc^2
\frac\alpha 2.
\end{eqnarray*}

\noindent
Note \cite{H} that if $c$ is a root length, then, $R_c$ is nothing but
the $W$-orbit of $\alpha$.

The Olshanetsky-Perelomov Hamiltonians with trigonometric potentials
associated to a root system $R$ are defined in terms of the above
data by
$$
\Hop=-\Delta + \sum_c a_c U_c,
$$
where the sum is taken over all root lengths, $c$, and where the
$a_c$'s are real coupling constants.

\section{The Algebraic Exact Solvability of $\Hop$.}

The affine hyperplanes $\{\lambda\in \Vsp^*:(\alpha,\lambda)\in
2\pi\integers\}$ determine in $\Vsp^*$ a set of isometric open bounded
subsets called alcoves.  Let $A$ denote the unique alcove (usually
referred to as the fundamental alcove) that is contained in $C$ and
that has the origin as a boundary point. Let $m$ denote the Lebesgue
measure on $A$.  From now on we use the inner product to identify $A$
with the corresponding subset of $\Vsp$ and restrict the domain of
functions introduced subsequently to $A$.  Our goal is to construct a
basis for the underlying Hilbert space $L^2(A,m)$ in which the
algebraic exact solvability of $\Hop$ is manifest. The elements of
this basis will be products of $W$-invariant trigonometric functions
of certain linear forms on $\Vsp$ with a common gauge factor vanishing
along the walls $\{u\in \Vsp:\alpha(u) \in 2\pi\integers\},\,
\alpha\in R$ of the potential terms $U_c$.

We now proceed to define this basis. Recall that a choice of positive
roots naturally induces a partial order relation, $\leq$, on the
weight lattice.  For $\lambda\in P^+$ set
\begin{align*}
  P_\lambda &= \bigcup_{w\in W} \{w(\mu) : \mu\in P^+ \text{ and }
  \mu \leq \lambda\},\\
  P_{\lambda^-} &= \bigcup_{w\in W} \{w(\mu) : \mu\in P^+ \text{ and }
  \mu \lneqq \lambda\},
\end{align*}

For $S \subset \Vsp^*$ let $\trig(S)$ denote the complex vector space
spanned by functions of the form $\E^{\I\lambda},\, \lambda\in S$. If
$S$ is a $W$-invariant subset of $\Vsp^*$, then there is a well
defined action of $W$ on $\trig(S)$, namely
$$w\cdot \E^{\I\lambda} = \E^{\I w(\lambda)},\quad w\in W,
\lambda\in S.$$
In this case, let $\trig(S)^W$ denote the subspace of
$W$-invariant functions.

Recall that a root system $R$ is said to be reduced if for every
$\alpha\in R$, the only roots homothetic to $\alpha$ are $-\alpha$ and
$\alpha$ itself.  A root $\alpha$ will be called non-divisible if
$\alpha/2$ is not a root.  Similarly, $\alpha$ will be called
non-multiplicable if $2\alpha$ is not a root.  Of course, if $R$ is
reduced, then all roots are both non-divisible and non-multiplicable.
An irreducible non-reduced system must be isomorphic to a root system
of type $\mathrm{BC}_n$ for some $n$.  To describe the latter, take
$\Vsp=\reals^n$ and let $\epsilon_1,\ldots, \epsilon_n$ denote the
dual basis of the standard basis of $\reals^n$.  The root system in
question consists of three types of roots: short roots $\pm
\epsilon_i$, medium roots $\pm \epsilon_i\pm \epsilon_j,\, i\neq j$ ,
and long roots $\pm 2\epsilon_i$.

For reasons which will become clear later, it is convenient to re-express the
coupling constants $a_c$ appearing in $\Hop$ as follows. We let $a_c=
k_c(k_c-1)$ if
$c$ is the length of an non-multiplicable root, and $a_c = k_c(k_c+k_{2c}-1)$
if
$R$ is non-reduced and $c$ is the length of the short roots. 
Let 
$$
A_c= \prod_{\alpha\in R_c^+} \sin\frac\alpha2 ,\quad
F=\prod_c |A_c|^{k_c},\quad
\rho_c= \frac12 \sum_{\alpha\in R_c^+} \alpha,\quad
\rho =\sum_c k_c\rho_c.
$$

The following theorems, which are the main results of our paper, shows
that the Olshanetsky-Perelomov trigonometric Hamiltonians $\Hop$ are
exactly solvable in the algebraic sense, and that the corresponding
eigenfunctions are physically meaningful.

\begin{theorem}
\label{thrm:es}
  Let $\lambda$ be a dominant weight .  If
$k_c\geq 0$ for each root
  length $c$, then there exists a unique
  $\phi_\lambda\in\trig(P_\lambda)^W$ such that $F\phi_\lambda$ is an
  eigenfunction of $\Hop$ with eigenvalue $\|\lambda+\rho\|^2$.
  Furthermore, if $F\phi,\, \phi\in\trig(P)^W$ is an eigenfunction of
  $\Hop$, then $\phi=\phi_\lambda$ for some $\lambda\in P^+$.
\end{theorem}

\begin{theorem}
\label {thrm:compl}
The subspace $F\trig(P)^W$ is dense in $L^2(A,m)$. Moreover, if
$k_c\geq0$ for all root lengths $c$, then the operator $\Hop$ is
essentially self-adjoint on the domain $F\trig(P)^W\subset L^2(A,m)$.
\end{theorem}

We begin with the proof of Theorem 2, assuming Theorem 1 to be true. We first have:

\begin{lemma}
  Let $D$ be an open, bounded subset of Euclidean space , and
  $f:D\rightarrow\reals$ a bounded continuous function that does
  not vanish on $D$ {\rm (} but may vanish on the boundary {\rm)}.
  With these assumptions, $f L^2(D,m)$ is a dense subset of
  $L^2(D,m)$.
\end{lemma}
\begin{proof}
  Let $D_0$, an open subset of $D$, be given, and choose
  $D_1$ such that $\overline{D}_1 \subset {D}_0$ and such that
  $m(D_0)-m(D_1)$ is smaller than a given $\epsilon>0$.  Note
  that $h=f^{-1}\chi_{ D_1}$ is a well defined
  element of $L^2(D)$ and that
  $fh=\chi_{D_1}$.  Consequently $\chi_{D_0}$
  lies in the closure of $fL^2(D)$.  The  conclusion follows
  from the fact that the characteristic functions form a dense
  subset of $L^2(D)$.
\end{proof}

\begin{proof}[Proof of Theorem \ref{thrm:compl}]
  Let $\Tsp$ denote the torus $\Vsp^*/{(2{\pi} Q)}$. We use
  the inner product on $\Vsp$ to identify $\Tsp$ with the identical quotient
  of $\Vsp$.  Recall that $\trig(P)$ is dense in $L^2(\Tsp)$ by the
  Fourier representation theorem.  Now $W$ acts on $\Tsp$ and $A$
  serves as a fundamental region for this action \cite[Ch. VI, \numero
  2.1]{Bourbaki}.  Consequently $\trig(P)^W$ is dense in $L^2(\Tsp)^W$
  and the latter is naturally isomorphic to $L^2(A,m)$.  We therefore
  conclude that $F\trig(P)^W$ is dense in $L^2(A,m)$ by applying the
  preceding Lemma with $f=F$. We now prove the essential
  self-adjointness of $\Hop$ on the domain $F\trig(P)^W$.  Let
  $A_0\subset A$ be an open subset with a piece-wise smooth boundary.
  Let $\phi_1,\phi_2\in\trig(P)^W$ be given.  Setting
  $\psi_i=F\phi_i,\, i=1,2$, we have
  \begin{align*}
    \int_{A_0} \Hop(\psi_1)\psi_2 - \psi_1 \Hop(\psi_2) &=
    \int_{A_0} \mathop{\mathrm{div}}(\psi_2 \nabla\psi_1 -
    \psi_1 \nabla\psi_2) \\
    &= \int_{\partial A_0} F^2(\phi_2 \nabla\phi_1 -\phi_1
    \nabla\phi_2)
  \end{align*}
  Hence, as the boundary of $A_0$ approaches the boundary of
  $A$, the above integrals tend to zero, so that the operator $\Hop$ is a
  symmetric.  By Theorem \ref{thrm:es} and the density of $F\trig(P)^W$ in
$L^2(A,m)$  , the span of eigenfunctions of
$\Hop$ is dense in
  $L^2(A)$, and therefore $\Hop$ must be essentially self-adjoint.
\end{proof}

 We now proceed with the proof of Theorem 1. The strategy behind the proof of this
theorem is to conjugate the Olshanetsky-Perelomov Hamiltonians $\Hop$ by a suitable
multiplication operator chosen in such a way that the resulting operator has a simple
action on the space $\trig(P)^W$. This will give rise to an essential intertwining
relation which will in turn imply the algebraic exact solvability. In order to
determine this multiplicative factor, we need a series of facts about
root lengths.

Let $M_c:W\rightarrow\{\pm 1\}$ be the class function defined by
$$M_c(s_\alpha)=
\begin{cases}
-1 & \text{if } \alpha\in B\cap R_c \\
\,1 & \text{if } \alpha\in B\backslash R_c
\end{cases}
$$
The following result is a straightforward consequence of the definition of
$A_c$:
\begin{proposition}
  For $w\in W$ one has $w(A_c) = M_c(w) A_c$.  In other words,
  $A_c$ is a relative invariant of $W$ with multiplier $M_c$.
\end{proposition}

\noindent Moreover, we have:

\begin{proposition}
  \label{prop:alpha-rho}
  Let $c$ be a root length.  If $\alpha\in B$, then $(
  \ckalpha,\rho_c )$ takes one of four possible values: $1$ if
  $\|\alpha\|=c$, $2$ if $\|\alpha\|=c/2$, $1/2$ if $\|\alpha\|=2c$,
  $0$ in all other cases.
\end{proposition}
\begin{proof}
  Let $\alpha\in B$ be given.  The action of $s_\alpha$ maps
  $\alpha$ to $-\alpha$ and permutes the elements of $R^+$ not
  homothetic to $\alpha$ \cite[Ch. VI, \numero 1.6]{Bourbaki}.  Let
  $\beta\in R_c^+$ be given and set $\beta'=s_\alpha(\beta)$.  Note
  that if $\beta=\beta'$, then $( \ckalpha , \beta) = 0$; and that if
  $\beta'\neq \beta$, then $( \ckalpha , \beta+\beta') = 0$. If
  $\|\alpha\|\notin\{c,2c,c/2\}$, then $\alpha$ is not homothetic to
  any element of $R_c$, and hence one can break up $\rho_c$ into
  subterms of length one and two such that each subterm is annihilated
  by $\ckalpha$.  This proves the fourth assertion of the
  proposition.  If $\|\alpha\|=c$, then $\rho_c$ is the sum of
  $\alpha/2$ and a remainder perpendicular to $\ckalpha$.
  Consequently, $( \ckalpha , \rho_c ) = 1,$ thereby proving the first assertion.  If
$\|\alpha\| = c/2$, then $2\alpha$ is also a
  root, and consequently $\rho_c$ is the sum of $\alpha$ and a
  remainder perpendicular to $\ckalpha$.  This implies the second assertion. The case
three assertion is proven similarly.
\end{proof}

\begin{corollary}
  If $c$ is the length of an non-multiplicable root, then $\rho_c$ is a
  weight.  If $R$ is non-reduced, and $c$ is the length of the short
  roots, then $\rho_c$ is merely a half-weight.
\end{corollary}

\begin{corollary}
  \label{cor:alpha-rho-int}
  Let $c$ be a root length.  Then for all $\alpha\in R_c$, one has
  $( \ckalpha, \rho_c) \in \integers$.  
\end{corollary}
\begin{proof}
  If $c$ is the length of an non-multiplicable root, then the claim follows from 
preceding corollary.  Suppose then that
  $2c$ is also a root length.  For $\alpha\in R_c$ note that $2
  (2\alpha)\ck{} = \ckalpha$ and that $2\rho_c = \rho_{2c}$. Hence
  $$( \ckalpha, \rho_c) = ( (2\alpha)\ck{},
  \rho_{2c}).$$
  Since $2\alpha$ is non-multiplicable, the right hand
  side is an integer by the preceding corollary.
\end{proof}
\begin{corollary}
  \label{cor:rhoc2}
  Let $c$ be a root length and $w\in W$.  Then, $w(\rho_c)\in
  Q-\rho_c$
\end{corollary}
\begin{proof}
  Note that
  $$w(\rho_c) = \frac12 \sum_{\alpha\in R_c^+} \sigma_\alpha(w)
  \alpha,$$
  where $\sigma_\alpha(w)$ is either $1$ or $-1$.  Hence,
  $\rho_c+w(\rho_c)$ is the sum of all $\alpha\in R_c^+$ such that
  $\sigma_\alpha(w)=1$.
\end{proof}
 
We are now ready for the next step leading to the required
intertwining relation, which is to show that $\trig(P_\lambda)^W$
is an invariant subspace of $\nabla \log |A_c|$. First, we have:

\begin{proposition}
  \label{prop:relibase}
  Let $c$ be a root length. If $\phi\in\trig(P-\rho_c)$ is a relative
  invariant of $W$ with multiplier $M_c$, then $\phi=A_c\, \phi_0$ for
  some $\phi_0\in\trig(P)^W$.
\end{proposition}
\begin{proof}
  By assumption, $\phi_1 = \E^{\I\rho_c} \phi$ is an element of
  $\trig(P)$.  Let $\alpha\in R_c^+$ be given.  The first claim is
  that $\phi_1$ is divisible by $\E^{\I\alpha}-1$ in $\trig(P)$.  By
  assumption, $\phi$ is a linear combination of expressions of the
  form $\E^{\I\lambda} - \E^{\I \lambda'}$ where $\lambda+\rho_c\in
  P$, and $\lambda'=s_\alpha(\lambda)$.  Since $\lambda$ is the
  difference of a weight and $\rho_c$, Corollary
  \ref{cor:alpha-rho-int} shows that $( \ckalpha , \lambda)
  \in \integers$. By switching $\lambda$ and $\lambda'$, if necessary,
  one may assume without loss of generality that $-(
  \ckalpha,\lambda) \in \natnums $.  The claim follows by noting
  that
  $$
  \E^{\I\lambda} - \E^{\I\lambda'} = \E^{\I\lambda} \lp 1-\E^{-\I(
    \ckalpha, \lambda ) \alpha} \rp,
  $$
  and by factoring the right hand side in the usual fashion.
  
  Note that $\trig(P)$ with the natural function multiplication is
  a unique factorization domain \cite[Ch. VI, \numero 3.1]{Bourbaki}.
  Hence the preceding claim implies that there exists a $\phi_0\in
  \trig(P)$ such that
  $$\phi_1 = \phi_0\prod_{\alpha\in R_c^+} \lp \E^{\I \alpha} - 1\rp.$$
  The proof is concluded by noting that up to a constant factor, $A_c$
  is equal to
  $$\E^{-\I\rho_c}\prod_{\alpha\in R_c^+} \lp \E^{\I \alpha} - 1\rp.$$
  The $W$-invariance of $\phi_0$ follows from the fact that $A_c$ and
  $\phi$ are relative invariants with the same multiplier.
\end{proof}

\noindent We have:

\begin{corollary}
  \label{cor:acformula}
  Let $c$ be a root length.  One has
  \begin{equation}
    \label{eq:ac}
    (2\I)^{\#R_c} A_c = \frac{1}{\# W_{\rho_c}} \sum_{w\in W} M_c(w)
    \E^{\I w( \rho_c)}. 
  \end{equation}
\end{corollary}

\begin{proposition}
  \label{prop:gradlog-action}
  The differential operator $\nabla \log |A_c|$ has a well-defined
  action on $\trig(P)^W$.
\end{proposition}
\begin{proof}
  Let $\phi\in \trig(P)^W$.  The claim is that $(\nabla \log
  |A_c|)\,(\phi)\in \trig(P)^W$.  By Corollaries \ref{cor:rhoc2} and
  \ref{cor:acformula}, $A_c \in \trig(Q-\rho_c)$, and hence $\nabla
  A_c(\phi)\in \trig(P-\rho_c)$.  Since $\nabla$ is a $W$-invariant
  operator, $\nabla A_c(\phi)$ is a relative invariant of $W$ with
  multiplier $M_c$.  Hence, by Proposition \ref{prop:relibase},
  there exists a $\phi_0\in \trig(P)^W$ such that $\nabla A_c(\phi) =
  A_c\phi_0$.
\end{proof}
We now have:
\begin{proposition}
  \label{prop:invspace}
  If $\lambda\in P^+$, then $\trig(P_\lambda)^W$ is an invariant
  subspace of $\nabla
  \log|A_c|$.
\end{proposition}
\begin{proof}
  Let $\phi\in\trig(P_\lambda)^W$ be given. Set $\phi_0 = (\nabla
  \log|A_c|)(\phi)$.  By Proposition \ref{prop:gradlog-action},
  $\phi_0\in\trig(P)^W$.  Let $\mu$ be a maximal element of
  $\supp(\phi_0)$.  Consequently $\mu+\rho_c$ is a maximal
  element of $\supp(A_c\phi_0)$. Now
  \begin{eqnarray*}
    A_c &=& b_1 \E^{\I\rho_c} + \text{ lower order terms } , \\
    \phi &=& b_2 \E^{\I\lambda} + \text{ lower order terms }, 
  \end{eqnarray*}
  where $b_1, b_2$ are non-zero constants, and hence,
  $$(\nabla A_c) (\phi) = -b_1 b_2 (\rho_c,\lambda) \E^{\I
    (\rho_c+\lambda)} + \text{ lower order terms }.
  $$
  Since $(\rho_c,\lambda)>0$, one must have $\rho_c+\lambda =
  \rho_c+\mu$.  Therefore $\mu=\lambda$, and
  $\phi_0\in\trig(P_\lambda)^W$.
\end{proof}

\noindent
The basic identity which will give rise to the intertwining relation which
we are looking for is given in the following proposition:
\begin{proposition}
  \label{prop:gauge-xform}
  Let $f_1,\ldots, f_n$ be smooth real-valued functions on $\Vsp$, let
  $k_1,\ldots, k_n$ be real constants and let
$$      X = \sum_{i=1}^n 2k_i \nabla \log |f_i|,\qquad
       F = \prod_{i=1}^n  |f_i|^{k_i}. 
$$
We have the identity
$$ F (-\Delta - X)  = (-\Delta + U) F,$$
where
$$
U = \sum_i k_i(k_i-1) \frac{\|\nabla f_i\|^2}{f_i^2} + \sum_{i\neq
  j} k_i k_j \frac{(\nabla f_i \, , \nabla f_j)}{f_i f_j}+\sum_i k_i
\frac{\Delta f_i}{f_i}.
$$
\end{proposition}

The application of this proposition to the Olshanetsky-Perelomov
Hamiltonians $\Hop$ requires a number of intermediate formulas.

\begin{proposition}
  \label{prop:nabla-ac}
  Let $c$ be a root length.  One has
  \begin{align}
    \label{eq:delta_ac}
    \Delta A_c &= -\|\rho_c\|^2 A_c ,\\
    \label{eq:nabla_ac2}
    \| \nabla A_c \|^2 &= \lp U_c - \| \rho_c \|^2 \rp A_c^2 , 
   \end{align}
\end{proposition}
\begin{proof}
  Note that for $\lambda\in \Vsp^*$ one has $\Delta \E^{\I\lambda} = -
  \|\lambda\|^2 \E^{\I\lambda}.$ Formula \eqref{eq:delta_ac} follows
  immediately from \eqref{eq:ac}.  Note that
  \begin{equation}
    \label{eq:nabla_ac}
    \nabla A_c = \frac{A_c}2 \sum_{\alpha\in R_c^+}
    \cot\frac\alpha2 \nabla\alpha.  
  \end{equation}
  Consequently,
  \begin{equation}
    \label{eq:nabac1} 
    \|\nabla A_c\|^2 = 
    \lp \frac{c^2}4 \sum_\alpha \cot^2 \frac\alpha 2
        + \frac14 \sum_{\alpha\neq\beta} (\alpha,\beta)\,
              \cot\frac\alpha2\,\cot\frac\beta2
    \rp A_c^2
  \end{equation}
  Taking the divergence of \eqref{eq:nabla_ac} one obtains
  $$
  \frac{\Delta A_c}{A_c} = -\frac{(\#R_c)\, c^2 }4 + \frac14
  \sum_{\alpha\neq\beta} (\alpha,\beta)\,
  \cot\frac\alpha2\,\cot\frac\beta2.
  $$
  Solving for the second term of
  the right hand side of the latter equation, substituting into \eqref{eq:nabac1}
and applying
  \eqref{eq:delta_ac}, we obtain \eqref{eq:nabla_ac2}.
  \end{proof}

\begin{proposition}
  \label{prop:nabla-aca2c}
  If $c_1$ , $c_2$ are distinct root lengths such that the
  corresponding roots are not  homothetic, then
  \begin{equation}
    \label{eq:nabla_ac1ac2}
    (\nabla A_{c_1}\,, \nabla A_{c_2}) = -(\rho_{c_1},\rho_{c_2}) \,A_{c_1}
    A_{c_2}
  \end{equation}
  If $R$ is non-reduced and $c$ is the length of the short roots, then
  \begin{equation}
    \label{eq:nabla_aca2c}
    (\nabla A_c\,, \nabla A_{2c}) = [\, U_c - (\rho_c,\rho_{2c})\, ] \,A_c
    A_{2c}
  \end{equation}
\end{proposition}
\begin{proof}
  Let $c_1$, $c_2$ be given.  A straightforward generalization of the
  argument in Proposition \ref{prop:relibase} yields
  $$
  A_{c_1} A_{c_2} = \frac1{\# W_{\rho_{c_1}+\rho_{c_2}}}
  \sum_{w\in W} M_{c_1}\!(w)\, M_{c_2}\!(w)\, \E^{\I
    w(\rho_{c_1}+\rho_{c_2})}.
  $$
  Hence,
  $$
  \Delta (A_{c_1} A_{c_2}) = -\| \rho_{c_1}+\rho_{c_2}\|^2 A_{c_1} A_{c_2},
  $$
  and the desired conclusion follows immediately from the usual
  product rule for the Laplacian.
  
  Next, assume that the second of the Proposition's hypotheses holds.
  Set $S_c=\prod_{\alpha\in R_c^+} \cos(\alpha/2),$ and note that
  $A_{2c} = 2A_c S_c.$ Since $R$ is of type $\mathrm{BC}_n$, a
  direct calculation will show that $\Delta S_c = -\|\rho_c\|^2 S_c$.
  Consequently,
  \begin{align*}
    2\,(\nabla A_c \,, \nabla S_c) &= \frac12\Delta A_{2c} - A_c
    \Delta
    S_c - S_c\Delta A_c = -\|\rho_c\|^2 A_{2c}. \\
    (\nabla A_c\,, \nabla A_{2c}) &= -\|\rho_c\|^2 A_{2c} + 2 S_c\,
    \|\nabla A_c\|^2.
  \end{align*}
  The formula to be proved now follows from \eqref{eq:nabla_ac2}.
\end{proof}

We can now state and prove the intertwining relation which is 
fundamental to the proof of our main result.
\begin{proposition}
  \label{prop:hamgauge}
 Let
 $$
 \tHop = -\Delta - \sum_c 2k_c \nabla \log |A_c|. 
 $$
 We have
  $$
  F\tHop = \Hop F -\|\rho\|^2. $$
\end{proposition}
\begin{proof}
  Apply Propositions \ref{prop:gauge-xform}, \ref{prop:nabla-ac},
  \ref{prop:nabla-aca2c}.
\end{proof}

Finally, we are ready to give the proof of Theorem 1, that is of the
algebraic exact solvability of the Olshanetsky-Perelomov Hamiltonian
$\Hop$.  We begin with the following simple result from linear algebra.
\begin{proposition}
  \label{prop:codim1}
  Let $\Vsp$ a finite-dimensional vector space over $\cnums$, and
  $\Vsp_1\subset \Vsp$ a codimension $1$ subspace. Let $T$ be an
  endomorphism of $\Vsp$ such that $\Vsp_1$ is an invariant subspace, and
  let $\kappa\in\cnums$ denote the unique eigenvalue of the
  corresponding endomorphism of $\Vsp/\Vsp_1$.  If $\kappa$ is not an
  eigenvalue of $T|_{\Vsp_1}$, then $\kappa$ is a multiplicity $1$
  eigenvalue of $T$.
\end{proposition}

It should be noted that the assumption $k_c\geq0$ in Theorem
\ref{thrm:es} is crucial.  The necessity of this assumption is
explained by the following Proposition. Indeed, one should remark that
there exist certain negative values of $k_c$ for which the action of
$\Hop$ fails to be diagonalizable.
\begin{proposition}
  \label{prop:unique-evalue}
  Let $\mu<\lambda$ be dominant weights. If  $k_c\geq0$ for each root
  length $c$, then $\|\lambda+\rho\| > \|\mu+\rho\|$.
\end{proposition}
\begin{proof}
  Note that 
  $$
  \|\lambda+\rho\|^2 - \|\mu+\rho\|^2 = \|\lambda\|^2 - \|\mu\|^2
  + 2\,(\lambda-\mu,\rho).
  $$
  Using the fact that $\lambda-\mu\in P^+$ one can easily show
  that $\|\lambda\|>\|\mu\|$.  Furthermore, since $\lambda-\mu$ is a
  linear combination of basic roots with positive coefficients,
  Proposition \ref{prop:alpha-rho} implies that
  $(\lambda-\mu,\rho)>0$.
\end{proof}

Finally, we have:

\noindent 
\begin{proof}[Proof of Theorem \ref{thrm:es}]
  
  Let $\lambda$ be a dominant weight. By Proposition
  \ref{prop:invspace}, $\trig(P_\lambda)^W$ is an invariant
  subspace of $\tHop$.  Using an argument similar to the one given in
  the proof of
  Proposition \ref{prop:invspace}, it is not hard to verify that if %
  $\phi\in\trig(P_\lambda)^W$, then
  \begin{equation}
    \label{eq:Hcod1}
    \lp \tHop - \|\lambda\|^2 - 2\,(\rho,\lambda) \rp
    (\phi)\in\trig(P_{\lambda^-})^W. 
  \end{equation}
  Note that $\trig(P_{\lambda^-})^W$ is a codimension $1$ subspace
  of $\trig(P_\lambda)^W$.  Furthermore, by Proposition
  \ref{prop:unique-evalue},
  $$
  \|\lambda\|^2 + 2(\lambda,\rho) > \|\mu\|^2+2(\mu,\rho)$$
  for all
  dominant weights $\mu<\lambda$. Hence, by Proposition
  \ref{prop:codim1}, there exists a unique
  $\phi_\lambda\in\trig(P_\lambda)^W$ such that $\tHop\phi_\lambda =
  (\|\lambda\|^2 + 2(\rho,\lambda))\phi$.  The first of the desired
  conclusions now follows by Proposition \ref{prop:hamgauge}.
  
  To prove the converse let $F\phi$ with $\phi\in\trig(P)^W$ be an
  eigenfunction of $\Hop$ with eigenvalue $\kappa$.  Let $\lambda\in
  P^+$ be a maximal element of $\supp(\phi)$.  Since
  $\trig(P_{\lambda^-})^W$ is a codimension $1$ subspace of
  $\trig(P_\lambda)^W$, \eqref{eq:Hcod1} implies that
  $\kappa=\|\lambda\|^2+2(\lambda,\rho)$.  Consequently $\lambda$ is
  the unique maximal element of $\supp(\phi)$.  By Proposition
  \ref{prop:codim1}, $\kappa$ has multiplicity $1$, and this gives the
  desired conclusion.
\end{proof} 
%One should remark that if vector negation is an element of $W$, then
%elements of $\trig(P)^W$ are real valued functions.  Since the gauge
%factor $F$ is also real valued, the same can be said regarding the
%eigenfunctions of $\Hop$.

\section{A recursion formula for the eigenfunctions of $\tHop$}
\label{sect:recform}
In the present section we show how to explicitly compute the
eigenfunctions of the Olshanetsky-Perelomov Hamiltonian by using a
$k_c$-para\-meterized analogue of the Freudenthal multiplicity
formula.  The generalized formula actually yields the eigenfunctions
$\phi_\lambda$ of the related operator $\tHop$.  One should mention
that the eigenfunctions $\phi_\lambda$ first appeared in the
investigations of Heckman and Opdam \cite{HO}, who regard these functions
as multi-variable generalizations of the Jacobi polynomials.  The
eigenfunctions of $\Hop$ are of course obtained by multiplication with
the gauge factor $F$.

By way of motivation it will be useful to recall the context of the
original Freudenthal formula.  Suppose that $R$ is reduced and let
$\chi_\lambda,\,\lambda\in P^+$ denote a character of the
corresponding compact, simply connected Lie group.  The Weyl character
formula states that
\begin{equation}
  \label{eq:wcf}
\chi_\lambda = 
\frac{\sum_{w\in W }\sgn(w)\, \E^{\I  w(\lambda+\trho)}}
{\sum_{w \in W }\sgn(w)\, \E^{\I  w(\lambda)}}
\end{equation}
where $\trho$ is the half-sum of the positive roots.  Now if $k_c=1$
for all $c$, then the potential term of $\Hop$ is zero, and the gauge
factor $F$ is nothing but the $W$-antisymmetric denominator of
\eqref{eq:wcf}.  Furthermore the numerator in \eqref{eq:wcf} is the
unique $W$-antisymmetric eigenfunction of $\Delta$ with highest order
term $\E^{\I(\lambda+\trho)}$. Hence, by the intertwining relation
described in Proposition \ref{prop:hamgauge}, the Weyl character
formula is equivalent to the statement that $\chi_\lambda$ is an
eigenfunction of $\tHop$ with eigenvalue $(\lambda,\lambda+2\trho)$.
This observation leads directly to the classical Freudenthal formula
for the multiplicities of $\chi_\lambda$, and to the following
generalization involving the parameters $k_c$. (See \cite{FH} for more
details regarding the Weyl and Freudenthal formulas.)
\begin{proposition}
  Let $\phi_\lambda=\E^{\I\lambda} + \sum_{\mu<\lambda} n_\mu
  \E^{\I\mu}$ be the eigenfunction of $\tHop$ described in the
  statement and proof of Theorem \ref{thrm:es}.  Setting $n_\lambda=1$
  and $n_\nu=0$ for $\nu\not\leq \lambda$, the remaining coefficients
  $n_\mu,\,\mu<\lambda$, are given by the following recursion formula:
  \begin{equation}
    \label{eq:fmf}
  (\|\lambda+\rho\|^2-\|\mu+\rho\|^2)\,n_\mu =
2\sum_{\alpha\in R^+}
\sum_{j\geq1}  k_{|\alpha|}\, (\alpha,\mu+j\alpha)\,n_{\mu+j\alpha}
  \end{equation}
\end{proposition}
\begin{proof}
  Rewriting 
  $$A_c=\E^{\I\rho_c}\prod_{\alpha\in R_c^+} \lp
  1-\E^{-\I\alpha}\rp,
  $$
  one obtains
  $$
  \tHop = -\Delta - \I\,\nabla\!\rho-2\,\I\!\sum_{\alpha\in R^+}
  k_{|\alpha|} \frac{\E^{-\I\alpha}}{1-\E^{-\I\alpha}}\nabla{\alpha}
  $$
  Let $\trig((P))$ denote the vector space of formal power
  series $\sum_{\mu\in P} c_\mu \E^{\I\mu}$.  Since elements of
  $\trig(P)$ are finitely supported sums, one has a well
  defined-multiplication operation $\trig((P))\times\trig(P)
  \rightarrow \trig((P)).$ Thus, setting the domain of $\tHop$ to
  be $\trig(P)$ one can extend the operator's coefficient ring and
  write
  $$
  \tHop = -\Delta - \I\,\nabla\!\rho-2\,\I\!\sum_{\alpha\in
    R^+}\sum_{j\geq 1} k_{|\alpha|}\, \E^{-j\I \alpha}
  \nabla{\alpha}
  $$
  However, because of Proposition \ref{prop:gradlog-action} one can
  take the codomain of $\tHop$ to be $\trig(P)$ rather than all of
  $\trig((P))$.  Acting with the right hand side of the latter equation on $\phi_\lambda$,
  collecting like terms, and using the fact that $\phi_\lambda$ is an
  eigenfunction with eigenvalue $(\lambda,\lambda+2\rho)$ immediately
  yields \eqref{eq:fmf}.
\end{proof}
It is important to remark that by Proposition \ref{prop:unique-evalue}
the coefficient of $n_\mu$ appearing in \eqref{eq:fmf} is never zero.
Consequently \eqref{eq:fmf} can indeed be used as a recursive formula for the
coefficients $n_\mu$.  One should also remark that the $W$-symmetry of
$\phi_\lambda$ means that it suffices to use formula \eqref{eq:fmf} to
calculate $n_\mu$ with $\mu\in P^+$.

\end{document}